\documentclass[twoside]{article}
\usepackage[a4paper,margin=1.75in]{geometry}

\usepackage[dvipsnames]{xcolor}
\usepackage{authblk}
\usepackage{graphicx,float,latexsym,amssymb,subeqnarray,amsmath,url,amsfonts,epstopdf,color,bbm,here}
\usepackage[numbers,square]{natbib}
\usepackage{pgfplots}
\usepackage{tikz}
\usepackage{tikzsymbols}
\usepackage{amsfonts}
\usepackage{dashrule}
\usepackage{amssymb}
\usepackage{enumerate}
\usepackage{subcaption}

\definecolor{colorMCA}{rgb}{1,0.5020,0.6706}
\definecolor{colorMLMC}{rgb}{0.3804,0.3804,0.3804}
\definecolor{colorSGBM}{rgb}{1,0.6706,0}
\definecolor{colormSABR}{rgb}{0.3922,1,0.8549}
\definecolor{colorFGL}{rgb}{0.8784,0.2510,0.9843}
\definecolor{colorADI}{rgb}{0.3922,0.8667,0.0902}
\definecolor{colorRBFFD}{rgb}{0.2667,0.5412,1}
\definecolor{colorRBFPUM}{rgb}{1,0.3216,0.3216}

\newlength{\lstyle}
\newlength{\symsp}

\newcommand{\dif}{\mathop{}\!\mathrm{d}}

\def\1{{\mathds{1}}}            % Indicator function

\title{A High Order Method for Pricing of Financial Derivatives using Radial Basis Function generated Finite Differences}
\author{Slobodan Milovanovi\'c and Lina von Sydow}
\affil{\small{Department of Information Technology \\ Uppsala University \\ Sweden}}

\date{}                     %% if you don't need date to appear

\begin{document}

\maketitle
\begin{abstract}
In this paper, we consider the numerical pricing of financial derivatives using Radial Basis Function generated Finite Differences in space. Such discretization methods have the advantage of not requiring Cartesian grids. Instead, the nodes can be placed with higher density in areas where there is a need for higher accuracy. Still, the discretization matrix is fairly sparse. As a model problem, we consider the pricing of European options in 2D. Since such options have a discontinuity in the first derivative of the payoff function which prohibits high order convergence, we smooth this function using an established technique for Cartesian grids. Numerical experiments show that we acquire a fourth order scheme in space, both for the uniform and the nonuniform node layouts that we use. The high order method with the nonuniform node layout achieves very high accuracy with relatively few nodes. This renders the potential for solving pricing problems in higher spatial dimensions since the computational memory and time demand become much smaller with this method compared to standard techniques.
\end{abstract}
%{\em{AMS Subject Classification:}} 

\section{Introduction}
\label{sec:intro}
In this paper, we are concerned with the numerical pricing of financial derivatives. A financial derivative is a contract whose value depends on an underlying asset such as a stock, an interest rate, or a commodity. The trading in financial derivatives has increased tremendously during the last decades, mostly due to the possibility to hedge positions in the underlying asset. Another important feature of financial derivatives is the potential for leverage, since a small movement in the underlying asset can cause a large movement in the value of the financial derivative. 

Due to the large traded volume of financial derivatives, efficient and accurate pricing of such contracts is of utmost importance. In most cases, there is no analytical formula available, and it becomes necessary to use a numerical method to compute the prices of the contracts. When a financial derivative depends on several underlying assets, the problem becomes multi-dimensional. Traditionally, the only way to price such financial derivatives is to use Monte Carlo methods for a stochastic differential equation (SDE) formulation of the problem. However, due to their arguably slow convergence, considerable efforts in the research community have been devoted to deriving efficient methods for a partial differential equation (PDE) formulation of the pricing problem. The main problem with these methods is the so-called curse of dimensionality --- the number of degrees of freedom in the problem grows exponentially in the number of dimensions. 

Numerical methods for the PDE formulation include adaptive Finite Differences (FD) \cite{persson2007pricing, lotstedt2007space, linde2009highly, persson2010pricing, von2015adaptive}, high-order compact schemes \cite{during2012high, during2015high}, Alternating Direction Implicit (ADI) schemes \cite{hout2010, ADI2}, Radial Basis Function (RBF) approximation \cite{pettersson2008improved, larsson2008multi}, Radial Basis Function Partition of Unity (RBF-PU) method \cite{safdari2015radial, shcherbakov2016radial,shcherbakov2016radialb}, and Radial Basis Function generated Finite Differences (RBF-FD) \cite{milovanovic2018radial,Slobodan}. In \cite{von2015benchop} and  \cite{benchop2} several methods for pricing of options are implemented and evaluated. 

As a numerical example, we consider pricing of a European two-dimensional option. An option is a financial derivative which gives the holder the right, but not the obligation, to buy (for call options) or sell (for put options) an underlying asset at a specified strike price $K$ at or before the time of maturity $T$. The method that we employ is RBF-FD. The main idea behind it is to combine the desirable features from FD (sparsity of the discretization matrices --- as opposed to RBF) and RBF (meshfree --- as opposed to FD). Such methods have the potential to be of high order, depending on the number of nodes used in the discretization stencil. However, for many option pricing problems, the payoff function has a discontinuity in the function itself or its derivatives, which limits the order of convergence obtained in numerical simulations. For this reason, we smooth the payoff function according to \cite{Kreiss}, before employing the numerical method. This smoothing increases the order of convergence to the expected one from the discretization that is used.

In Section \ref{sec:nummet} we define the discretization in space and time, while Section \ref{sec:M} is devoted to the model problems that we solve, as well as node layouts, stencils, boundary conditions, and smoothing of the initial data. Finally, the results are presented in Section \ref{sec:numres}, and conclusions are drawn in Section \ref{sec:concl}.

\section{Numerical method}
\label{sec:nummet}

We consider pricing of financial derivatives where the problem can be formulated as a PDE in $D$ spatial dimensions and time 
% To facilitate the use of standard time-stepping schemes, we transform (\ref{eq:L}) into a problem in forward time such that  
%\begin{equation}\label{eq:L2}
%\begin{array}{rcl}
%{\displaystyle{\frac{\partial u}{\partial t}+{\mathcal{L}}u}}&=&0,\\
%u(s_1,\ldots,s_D,0)&=&g(s_1,\ldots,s_D),\\
%s_i\ge 0,\,\,i=1,\ldots,D,&&0\le t\le T.
%\end{array}
%\end{equation}
\begin{align}
\frac{\partial u}{\partial t}+{\mathcal{L}}u&=0, \nonumber \\
u(s_1,\ldots,s_D,0)&=g(s_1,\ldots,s_D), \label{eq:L2} \\
s_i\ge 0,\ i=1,\ldots,D;& \quad 0\le t\le T. \nonumber
\end{align}
Here the solution $u(s_1,\ldots,s_D,t)$ denotes the price of the financial derivative, $t$ denotes time, $s_i$, the value of the underlying asset with index $i$, and $g$ the payoff function of the financial derivative. In many pricing problems the original PDE is a final value problem solved backward in time. We consider problems in forward time as in (\ref{eq:L2}), i.e., when necessary the problem is transformed into an initial value problem. 

In Sections \ref{sec:RBF-FD} and \ref{sec:temp} we define the spatial and temporal discretization of (\ref{eq:L2}), respectively.
%
%{\bf{\color{red}Kolla tecken}}
%
%\bigskip
%where 
%$$
%\mathcal{L}u=-r\sum\limits_{i}^{D}s_i\frac{\partial u}{\partial{s_i}}-\frac{1}{2}\sum\limits_{i,j}^{D}\rho_{i,j}\sigma_i\sigma_j s_is_j\frac{\partial^2u}{\partial s_i \partial s_j}ru.$$
%
%\bigskip
%{\color{red}\bf{Maybe something about American options.}}

\subsection{Radial Basis Function generated Finite Differences}
\label{sec:RBF-FD}

In RBF-FD the spatial operator $\mathcal{L}u$ in (\ref{eq:L2}) at a location  $\mathbf{s}^c=(s_1^c,s_2^c,\ldots,s_D^c)$, is approximated as a linear combination of the solution at the $m$ closest nodes $\mathbf{s}^k$ (possibly including $\mathbf{s}^c$), $k=1,\ldots,m$
\begin{equation}
\mathcal{L}u|_{\mathbf{s}^c}\approx\sum_{k=1}^{m}{w}_{k}u|_{\mathbf{s}^k}.
\label{eq:Lu}
\end{equation}
The weights ${w}_{k}$ are calculated by enforcing (\ref{eq:Lu}) to be exact for an RBF $\phi(r)$ 
\begin{equation}\label{eq:D}
{{
\begin{bmatrix}
\phi(\|\mathbf{s}^1-\mathbf{s}^1\|) & \ldots & \phi(\|\mathbf{s}^1-\mathbf{s}^m\|)\\
\vdots & \ddots & \vdots\\
\phi(\|\mathbf{s}^m-\mathbf{s}^1\|) & \ldots & \phi(\|\mathbf{s}^m-\mathbf{s}^m\|)
\end{bmatrix}
\begin{bmatrix}
{w}_1\\
\vdots\\
{w}_m
\end{bmatrix}=
\begin{bmatrix}
\mathcal{L}\phi(\|\mathbf{s}-\mathbf{s}^1\|)|_{{\mathbf{s}}^c}\\
\vdots \\
\mathcal{L}\phi(\|\mathbf{s}-\mathbf{s}^m\|)|_{{\mathbf{s}}^c}
\end{bmatrix}.}}
\end{equation}
It is given from RBF interpolation that (\ref{eq:D}) is a nonsingular system, and hence a unique set of weights  $w_k$, $k=1,\ldots,m$ can be computed.

%\begin{equation}
%\begin{bmatrix}
%\varphi(\|s_{1}^{(i)}-s_{1}^{(i)}\|) & \ldots & \varphi(\|s_{1}^{(i)}-s_{M}^{(i)}\|)\\
%\vdots & \ddots & \vdots\\
%\varphi(\|s_{M}^{(i)}-s_{1}^{(i)}\|) & \ldots & \varphi(\|s_{M}^{(i)}-s_{M}^{(i)}\|)
%\end{bmatrix}
%\begin{bmatrix}
%w_1\\
%\vdots\\
%w_M
%\end{bmatrix}=
%\begin{bmatrix}
%[{\cal{L}}\varphi(\|s-s_{1}^{(i)}\|)]_{s=s_i}\\
%\vdots\\
%[{\cal{L}}\varphi(\|s-s_{M}^{(i)}\|)]_{s=s_i}
%\end{bmatrix}.
%\label{eq:w}
%\end{equation}

Typical choices of RBFs are listed in Table \ref{tab:rbf}. For the first four examples, the parameter $\varepsilon\in \mathbb{R}$ is the shape parameter of the RBF. For polyharmonic splines (PHSs), the parameter $q\in{\mathbb{N}}$.

In this paper, we follow \cite{flyer2016role}, \cite{bayona2017role}, and \cite{Slobodan} and use PHSs as basis functions together with polynomials of degree $p$ in the interpolation. With that approach, the polynomial degree (instead of the RBF) controls the rate of convergence, while the RBFs contribute to reduction of approximation errors and are necessary in order to have a stable approximation. 
\begin{table}[H]
\centering
\begin{tabular}{|l|l|} \hline
\multicolumn{2}{|c|}{$\phi(r)$} \\ \hline
Gaussian&${\displaystyle{e^{-(\varepsilon r)^2}}}$\\
Inverse quadratic& ${\displaystyle{{1}/({1+(\varepsilon r)^2})}}$ \\
Multiquadric& ${\displaystyle{\sqrt{1+(\varepsilon r)^2}}}$ \\ 
Inverse multiquadric& ${\displaystyle{1/\sqrt{1+(\varepsilon r)^2}}}$ \\ \hline
Polyharmonic splines&$r^{2q-1}$ \\
\hline
\end{tabular}
\caption{A list of commonly used RBFs $\phi(r)$.}
\label{tab:rbf}
\end{table}

\noindent In (\ref{eq:Dc}), we augment (\ref{eq:D}) with monomials of degree one

\begin{equation}\label{eq:Dc}
{\small{\begin{bmatrix}
 & & &1&s^1_1&\ldots&s^1_D\\
& B&& 1&\vdots &&\vdots\\
 &  & &1&s^m_1&\ldots&s^m_D    \\
1&\ldots&1&0&0&\ldots&0\\
s^1_1&\ldots&s^m_1&0&0&\ldots&0\\
\vdots & \vdots & \vdots&\vdots &\vdots & \ddots&\vdots \\
s^1_D&\ldots&s^m_D&0&0&\ldots&0\\
\end{bmatrix}
\begin{bmatrix}
{w}_1\\
\vdots\\
{w}_m\\
\gamma_0\\
\gamma_1\\
\vdots\\
\gamma_D
\end{bmatrix}=
\begin{bmatrix}
\mathcal{L}\phi(\|\mathbf{s}-\mathbf{s}^1\|)|_{{\mathbf{s}}^c}\\
\vdots \\
\mathcal{L}\phi(\|\mathbf{s}-\mathbf{s}^m\|)|_{{\mathbf{s}}^c}\\
\mathcal{L}1|_{{\mathbf{s}}^c}\\
\mathcal{L}{s}_1|_{{\mathbf{s}}^c}\\
\vdots\\
\mathcal{L}{s}_D|_{{\mathbf{s}}^c}\\
\end{bmatrix},}}
\end{equation}
where $B$ is the coefficient-matrix in (\ref{eq:D}). 

%With polynomial augmentation of degree $p$ we get
%\begin{equation}\label{eq:Dp}
%{\small{\begin{bmatrix}
%B&P^T\\
%P&0
%\end{bmatrix}
%\begin{bmatrix}
%\bar{w} \\
%\bar{\gamma}
%\end{bmatrix}=
%\begin{bmatrix}
%\mathcal{L}\phi(\|\mathbf{s}-\mathbf{s}^1\|)|_{{\mathbf{s}}^c}\\
%\vdots \\
%\mathcal{L}\phi(\|\mathbf{s}-\mathbf{s}^m\|)|_{{\mathbf{s}}^c}\\
%\mathcal{L}1|_{{\mathbf{s}}^c}\\
%\mathcal{L}{s}_1|_{{\mathbf{s}}^c}\\
%\vdots\\
%\mathcal{L}{s}_D|_{{\mathbf{s}}^c}\\
%\end{bmatrix},}}
%\end{equation}

%\begin{align*}
%\begin{bmatrix}
%\varphi(\|s_{1}^{(i)}-s_{1}^{(i)}\|) & \ldots & \varphi(\|s_{1}^{(i)}-s_{M}^{(i)}\|)&1\\
%\vdots & \ddots & \vdots\\
%\varphi(\|s_{M}^{(i)}-s_{1}^{(i)}\|) & \ldots & \varphi(\|s_{M}^{(i)}-s_{M}^{(i)}\|)&1\\
%1&\ldots&1&0
%\end{bmatrix}
%\begin{bmatrix}
%w_1\\
%\vdots\\
%w_M\\
%\gamma
%\end{bmatrix}=
%\begin{bmatrix}
%[{\cal{L}}\varphi(\|s-s_{1}^{(i)}\|)]_{s=s_i}\\
%\vdots\\
%[{\cal{L}}\varphi(\|s-s_{M}^{(i)}\|)]_{s=s_i}\\
%[{\cal{L}}1]_{s=s_i}
%\end{bmatrix}.
%\end{align*}

Now, we place $N$ computational nodes $\mathbf{s}^c_i$, $i=1,\ldots,N$ at the locations where we want to approximate the solution. The weights for each computational node from solving (\ref{eq:Dc}) are assembled row-wise into the sparse differentiation matrix $W\in\mathbb{R}^{N\times N}$, with $m$ nonzero elements per row. This leads to the following semi-discretization of (\ref{eq:L2})
\begin{equation}
\label{eq:semi}
\begin{array}{rcl}
{\displaystyle{\frac{\dif }{\dif t}\bar{u}(t)+W \bar{u}(t)}}&=&\bar{0},\\
\bar{u}(0)&=&\bar{g},
\end{array}
\end{equation}
where $\bar{u}(t)\in\mathbb{R}^{N\times 1}$ is the vector of unknowns at time $t$, with approximations of $u$ in the computational nodes $\mathbf{s}^c_i$, $i=1,\ldots,N$, $\bar{0}\in\mathbb{R}^{N\times 1}$ is a vector with only zeros, and $\bar{g}\in\mathbb{R}^{N\times 1}$ is the vector with the function $g$ evaluated in the computational nodes $\mathbf{s}^c_i$, $i=1,\ldots,N$.
 
Equation (\ref{eq:semi}) forms a system of linear ordinary differential equations (ODEs) in time. In the next section, we describe how to solve it. 

\subsection{Temporal discretization}
\label{sec:temp}

For the time discretization of (\ref{eq:semi}), we use the Backward Differentiation Formula of order two (BDF2) \cite{Hairer2008}. This time-stepping scheme requires the solution at two previous time steps, and we therefore employ Backward Euler (BDF1) for the first time step. It is convenient to have the same coefficient matrix in all time steps, so we use non-equidistant time steps as described in \cite{larsson2008multi} and later used in e.g., \cite{milovanovic2018radial,Slobodan}. This is accomplished by discretizing the time interval with $M$ steps of length $\Delta t^{\ell} = t^{\ell} - t^{\ell-1}$, where $\ell = 1,\ldots,M$. We define $\omega_{\ell}=\Delta t^{\ell}/\Delta t^{\ell-1}$ for $\ell = 2,\ldots,M$ and arrive at
\begin{align}
\label{eq:BDF1} \bar{u}^1 - \bar{u}^0 &= \Delta t^1W\bar{u}^1,\\
\label{eq:BDF2} \bar{u}^{\ell} - \beta_1^\ell \bar{u}^{\ell-1} + \beta_2^\ell \bar{u}^{\ell-2} &= \beta_0^{\ell}W\bar{u}^{\ell},\quad \ell = 2,\ldots,M,
\end{align}
where
\begin{equation}
\beta_0^{\ell} = \Delta t^{\ell}\frac{1 + \omega_{\ell}}{1+ 2\omega_{\ell}},\quad \beta_1^{\ell} = \frac{(1+\omega_{\ell})^2}{1+2\omega_{\ell}},\quad \beta_2^{\ell} = \frac{\omega_{\ell}^2}{1+2\omega_{\ell}}.
\label{eq:betas}
\end{equation}
We compute the values for $\omega_{\ell}$ using the recursive condition $\beta_0^{\ell}=\beta_0^{\ell-1}$, which keeps the coefficient matrix constant throughout all time steps. Since our time interval has the length $T$, we chose the initial time step length $\Delta t^1$ from
\begin{equation}
\label{eq:Dt}
\sum_{\ell=1}^{M}\Delta t^{\ell} = T = \Delta t^1(1+\sum_{\ell=2}^{M} \prod_{\ell'=2}^{\ell}\omega^{\ell}).
\end{equation}
Finally, we start the time integration by setting $\bar{u}^0=\bar{g}$.

From the temporal discretization we get the following linear system of equations to solve in each time step
\begin{equation}
\label{eq:lineq}
A\bar{u}^{\ell}=\bar{b}^{\ell},
\end{equation}
where $A=I-\Delta t^1 W$, $\Delta t^1$ is given by (\ref{eq:Dt}), $\bar{b}^{\ell}=\beta_1^{\ell}\bar{u}^{\ell-1}-\beta_2^{\ell}\bar{u}^{\ell-2}$, $\ell=2,\ldots,M$, and $\bar{b}^1=\bar{u}^0=\bar{g}$. 
%We solve (\ref{eq:lineq}) using GMRES \cite{saad1986gmres} with an Incomplete LU-factorization (ILU) of $A$ as preconditioner.

\section{Model problem}
\label{sec:M}
%\subsection{PDE formulation}
%\label{sec:model}
As a model problem we consider a European call option issued on two underlying assets $s_1$ and $s_2$
\begin{align}
\frac{\partial u}{\partial t}+\mathcal{L}u&=0, \label{eq:Eur} \\
s_1 \ge 0,\ s_2 \ge 0,\ & \quad 0\le t \le T, \nonumber
\end{align}
%\begin{equation}\label{eq:Eur}
%\begin{array}{rcl}
%{\displaystyle{\frac{\partial u}{\partial t}+\mathcal{L}u}}&=&0,\\
%s_1\ge0,\,\,
%s_2\ge0,\,\,&&
%0\le t \le T,
%\end{array}
%\end{equation}
with
\begin{align}
\mathcal{L}u= \frac{1}{2}\Big(\sigma_1^2 s_1^2\frac{\partial^2u}{\partial s_1^2}&+\sigma_2^2 s_2^2\frac{\partial^2u}{\partial s_2^2}\Big)+ \rho\sigma_1\sigma_2 s_1s_2\frac{\partial^2u}{\partial s_1 \partial s_2} \nonumber \\
 &+ r\Big(s_1\frac{\partial u}{\partial{s_1}}+s_2\frac{\partial u}{\partial{s_2}}\Big) -ru, \label{eq:L}
\end{align}
%\begin{equation}
%\begin{array}{rcl}
%{\displaystyle{\mathcal{L}u}}&=&{\displaystyle{r\left(s_1\frac{\partial u}{\partial{s_1}}+s_2\frac{\partial u}{\partial{s_2}}\right)}}\\
%&+&{\displaystyle{\frac{1}{2}\left(\sigma_1^2 s_1^2\frac{\partial^2u}{\partial s_1^2}+\sigma_2^2 s_2^2\frac{\partial^2u}{\partial s_2^2}\right)}}\\
%&+&{\displaystyle{\rho\sigma_1\sigma_2 s_1s_2\frac{\partial^2u}{\partial s_1 \partial s_2}-ru}},\end{array}
%\label{eq:L}
%\end{equation}
and
\begin{equation}
u(s_1,s_2,T)=g(s_1,s_2)=\left(\frac{1}{2}(s_1+s_2)-K\right)^+.
\label{eq:init}
\end{equation}
Here $(f(x))^+=\max(f(x),0)$, $r$ denotes the risk-free interest rate in the market, $\sigma_i$ the volatility of asset $i$, and $\rho$ the correlation between the assets. As a close-field boundary condition in $s_1=s_2=0$ we set
\begin{equation}
u(0,0,t)=0,\quad 0\le t\le T,
\label{eq:close}
\end{equation}
and as a far-field boundary condition we set
\begin{equation}
u(s_1,s_2,t)=\left(\frac{1}{2}(s_1+s_2)-K e^{-rt}\right),\quad 0\le t\le T,
\label{eq:far}
\end{equation}
for $s_1$ and $s_2$ large enough.
%As a second model problem we consider a two-dimensional American put option
%\begin{equation}\label{eq:Am}
%\begin{array}{rcl}
%{\displaystyle{\frac{\partial u}{\partial t}+\mathcal{L}u}}&\ge&0,\\
%u(s_1,s_2,t)&\ge&g(s_1,s_2)\\
%{\displaystyle{\left(\frac{\partial u}{\partial t}+\mathcal{L}u\right)\left(u(s_1,s_2,t)-g(s_1,s_2) \right) }}&=&0,\\
%s_1\ge0,\,\,
%s_2\ge0\,\,&,&
%0\le t\le T.
%\end{array}
%\end{equation}
%with 
%\begin{equation}
%u(s_1,s_2,T)=g(s_1,s_2)=\left(K-\frac{1}{2}(s_1+s_2)\right)^+.
%\label{eq:initam}
%\end{equation}
%as final condition and $\mathcal{L}u$ defined by Equation (\ref{eq:L}). For this model problem we set the close-field boundary condition in $s_1=s_2=0$ to
%\begin{equation}
%\begin{array}{rcl}
%u(0,0,t)=K&,&0\le t\le T,
%\end{array}
%\label{eq:closeam}
%\end{equation}
%and as a far-field boundary condition we set
%\begin{equation}
%\begin{array}{rcl}
%u(s_1,s_2,t)=0&,&0\le t\le T,
%\end{array}
%\label{eq:faram}
%\end{equation}
%for $s_1$ and $s_2$ large enough.
The parameters used are given in Table \ref{tab:2}. 
{\color{red}
\begin{table}[htb]
\centering
\begin{tabular}{|c|c|} \hline
$r$&{\color{black}0.03} \\
$\sigma_1$&{\color{black}0.15} \\
$\sigma_2$& {\color{black}0.15}  \\
$\rho_{1,2}$& {\color{black}0.5} \\
$K$& {\color{black}1} \\
$T$& {\color{black}0.2}  \\
\hline
\end{tabular}
\caption{Parameters used in the model problem}
\label{tab:2}
\end{table}}

Equation (\ref{eq:Eur}) is a PDE that should be solved backward in time. To apply the time-stepping scheme in Section \ref{sec:temp}, we therefore transform (\ref{eq:Eur}) into a problem that is solved forward in time. 

\subsection{Node layout, stencils, and boundary conditions}
\label{sec:nodes}

\begin{figure}[H]
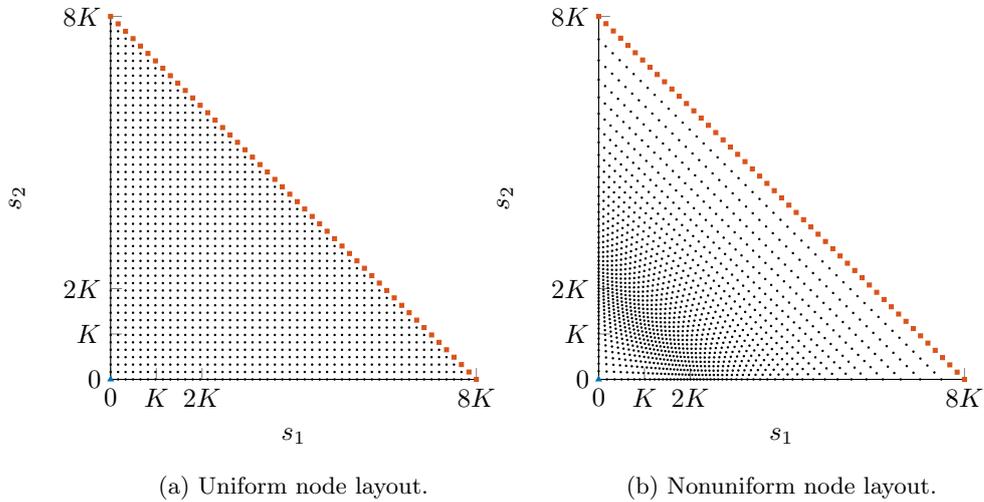

\centering
\makebox[\linewidth][c]{%
\centering
\begin{subfigure}[H]{0.8\textwidth}
\centering
\input{tikz/grid2dreg.tikz}
\caption{Uniform node layout.}
\end{subfigure} \hspace{-3.5cm}
\begin{subfigure}[H]{0.8\textwidth}
\centering
\input{tikz/grid2dadap.tikz}
\caption{Nonuniform node layout.}
\end{subfigure}
}
\caption{Uniform and nonuniform computational node layouts in 2D. The boundary conditions are employed in the blue triangle node (the close-field boundary condition) and in the red square nodes (the far-field boundary condition).}
\label{fig:221}
\end{figure}

We consider both a uniform and a nonuniform node layout, presented in Figure \ref{fig:221}. Unlike classical grid-based methods (e.g., standard FD methods) we do not need to use a rectangular domain. Instead, we only use the lower-triangular half of the rectangle which reduces the number of computational nodes by a factor of two, and hence the computational complexity significantly.

The reason for introducing a nonuniform node layout is that we can cluster nodes where we are most interested in having an accurate solution. In general, we are most interested in having an accurate solution in the neighborhood of $s_1+s_2=2K$, which is also where the truncation error is largest due to large derivatives in the solution from the discontinuity in the first derivative of the payoff function.

We start to present a nonuniform node distribution in 1D that is generated as introduced in \cite{hout2010} and later used for RBF-FD and option pricing in \cite{milovanovic2018radial}. Consider $N_1$ equidistant nodes $x_1^{(1)}<\ldots<x_1^{(i)}<\ldots<x_1^{(N_1)}$ constructed by
\begin{equation}
x_1^{(i)}=\text{arcsinh}\left(-\frac{K}{c}\right)+(i-1) \Delta x\text{,}\quad i=1,\ldots,N_1\text{,}
\label{eq:adap1}
\end{equation}
where $c$ is a positive real constant which specifies how dense the node distribution becomes around the strike price $K$,
$$\Delta x=\frac{1}{N_1}\left[\text{arcsinh}\left(\frac{s_{\max}-K}{c} \right) -  \text{arcsinh}\left( -\frac{K}{c}\right)\right]\text{,}$$
and $s_{\max}$ denotes the far-field boundary. Then, the nonuniform node distribution $s_1$ is generated pointwise as
\begin{equation}
s_1^{(i)}=K+c\cdot \text{sinh}(x^{(i)})\text{,}\quad i=1,\ldots,N_1\text{.}
\label{eq:adap2}
\end{equation}
%An example of a nonuniform node layout using this tuning parameter and $N=51$ is shown in Figure \ref{fig:21}\subref{fig:21b}.

The nonuniform node layout is generated by using the one-dimensional node layouts  from (\ref{eq:adap1}) and (\ref{eq:adap2}), along the axes $s_1$ and $s_2$, and then uniformly placing the internal points in the diagonal direction. The number of nodes along each diagonal is increased by one for each diagonal. The far-field boundary is located at $s_1+s_2=s_{\max}=8K$. The density tuning parameter used in Figure \ref{fig:221} for the nonuniform node layout and in the numerical experiments presented in Section \ref{sec:numres}, is $c=0.8$. It should be noted that a too small value of $c$ eventually leads to an ill-conditioned problem. 
%We also implemented uniform staggered node layouts, but numerical experiments show that they are equal or inferior to the non-staggered ones. 
We also introduce the notation $N_s$ for the number of nodes along one of the axes, i.e.,
\begin{equation}
\frac{N_s(N_s+1)}{2}=N.
\label{eq:Ns}
\end{equation}

\begin{figure}[H]
\centering
\input{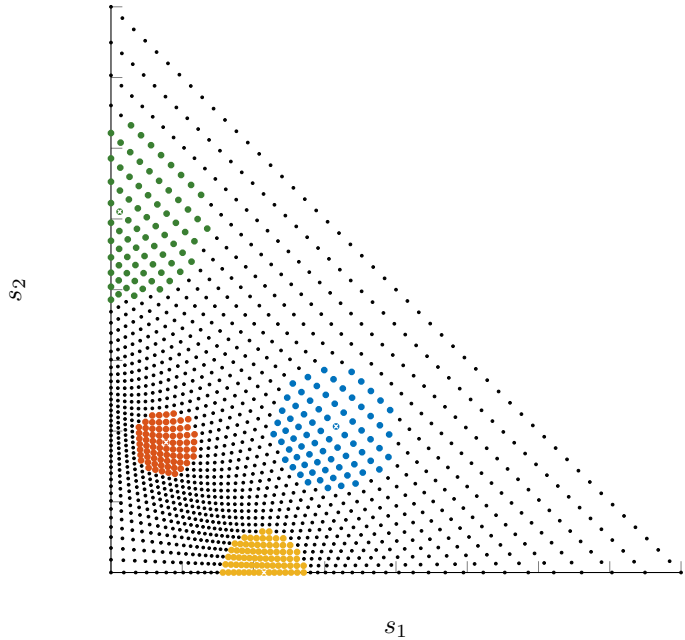}
\caption{Examples of nearest neighbor based stencils used for approximating the differential operator on a nonuniform node layout. The central node of each displayed stencil is denoted by a white cross mark. All stencils are of the same size $m=75$.}
\label{fig:222}
\end{figure}

The nearest neighbors for constructing the stencils are efficiently determined using the $k$-D tree algorithm, \cite{bentley1975multidimensional}. In Figure \ref{fig:222} we show examples of stencils at different locations in the domain. The polynomial space is of size 
$$\nu=\left(\begin{array}{c}
p+D\\
p\end{array}
\right),$$
which we use to set the size of the stencils to $m=5\nu$, following \cite{flyer2016role,bayona2017role,Slobodan}. We are aiming for a fourth order scheme and use $p=4$ and $q=5$ which gives $$\nu=\left(\begin{array}{c}
6\\
4\end{array}
\right)=15.$$ 
Hence, we use a stencil size that is $m=5\cdot 15=75$.

For the boundary nodes we use different treatments depending on where the node is located. For the node $s_1=s_2=0$ (the blue triangle in Figure \ref{fig:221}), we set the close-field boundary condition from (\ref{eq:close}). For the nodes $s_1+s_2=8K$ (the red squares in Figure \ref{fig:221}), we set the far-field boundary conditions from (\ref{eq:far}). For the boundary nodes along the axes, i.e., $s_1=0$, $s_2>0$ and $s_2=0$, $s_1>0$ we solve (\ref{eq:Eur}) using the discretization scheme defined in Section \ref{sec:nummet}. The $k$-D tree algorithm generates one-sided stencils for those nodes. 

\subsection{Smoothing of initial data}
\label{sec:smooth}
Since the initial data $g(s_1,s_2)$ in (\ref{eq:init}) has a discontinuity in the first derivative, the obtained spatial order of convergence for a finite difference scheme is limited to two, regardless of the formal order of the scheme. The formal spatial order of the scheme in (\ref{eq:Dc}) is $p$, i.e., for $p>2$ the obtained convergence order is limited by the lack of smoothness in the final condition. In \cite{Kreiss}, a smoothing of the initial data that recovers the order of convergence to the formal order of the scheme is introduced. This approach has been successfully used for option pricing problems in e.g., \cite{Pooley} and \cite{during2015high}.

Since we are aiming for a fourth order scheme, we use a fourth order smoothing operator $\Phi_4$ defined by its Fourier transform
\begin{equation}
\hat{\Phi}_4(\omega)=\left(\frac{\sin(\omega/2)}{\omega/2}\right)^4\left(1+\frac{2}{3}\sin^2(\omega/2)\right).
\label{eq:phihat}
\end{equation}
Using {\textsc{Wolfram Mathematica}} to compute the inverse Fourier transform of (\ref{eq:phihat}) gives
\begin{align}
{\Phi}_4(s) = \frac{1}{72}\Big(     &   -(s-3)^3 \cdot \mathrm{sgn}(s-3)        -(s+3)^3\cdot \mathrm{sgn}(s+3) \nonumber \\
                                                   &   +12(s-2)^3 \cdot \mathrm{sgn}(s-2)   +12(s+2)^3\cdot \mathrm{sgn}(s+2)  \nonumber \\
						&   -39 (s-1)^3 \cdot \mathrm{sgn}(s-1)   -39(s+1)^3\cdot \mathrm{sgn}(s+1) \nonumber \\
						& +56s^3 \cdot \mathrm{sgn}(s) \Big), \label{eq:phi}
\end{align}
%\begin{equation}
%\begin{array}{l}
%{\Phi}_4(s)={\textstyle{\left( - (s-3)^3\cdot \mathrm{sgn}(s-3) + 12(s-2)^3\cdot \mathrm{sgn}(s-2) -\right.}}  \\
%39(s-1)^3\cdot \mathrm{sgn}(s-1)
%+56s^3\cdot \mathrm{sgn}(s)-39(s+1)^3\cdot \mathrm{sgn(s+1)} + \\
%\left. 12(s+2)^3\cdot \mathrm{sgn}(s+2) - (s+3)^3\cdot \mathrm{sgn}(s+3)\right)/72, \nonumber
%\end{array}
%\label{eq:phi}
%\end{equation}
where
$$\mathrm{sgn}(x)=\frac{|x|}{x}.$$
Following \cite{Kreiss}, \cite{Pooley}, and \cite{during2015high}, we get the smoothed final condition on a uniform node layout as
\begin{equation}
{\displaystyle{
\tilde{g}(s_1,s_2)=\frac{1}{\Delta s^2}\int\limits_{-3\Delta s}^{3\Delta s}\int\limits_{-3\Delta s}^{3\Delta s}\Phi_4\left(\frac{\tilde{s}_1}{\Delta s}\right)\Phi_4\left(\frac{\tilde{s}_2}{\Delta s}\right)g(s_1-\tilde{s}_1,s_2-\tilde{s}_2)\dif\tilde{s}_1\dif\tilde{s}_2. }}
\label{eq:smooth}
\end{equation}
Since $g(s_1,s_2)$ is smooth in a large part of the computational domain, we only need to compute (\ref{eq:smooth}) in the nodes that are close enough to $s_1+s_2=2K$ to be affected from the smoothing. Also, since the nodes along a diagonal all have the same distance to $s_1+s_2=2K$, we only need to compute one value of $\tilde{g}(s_1,s_2)$ for each diagonal and use that value for all nodes on that diagonal.

The theory in \cite{Kreiss} shows that replacing the final condition $g(s_1,s_2)$ with $\tilde{g}(s_1,s_2)$ defined in (\ref{eq:smooth})  gives a fourth order scheme for Cartesian grids, i.e., the node layout that we here refer to as uniform. Here, we want to use this smoothing also for our nonuniform node layout defined in Section \ref{sec:nodes}. This layout can be seen as a slightly skewed Cartesian grid and the nodes are equidistantly distributed along the diagonals. For this node layout we replace $\Delta s$ in (\ref{eq:smooth}) with 
$$\Delta s_i=\min_{k = 1,\ldots,m}^{k \ne c} \|{\mathbf{s}}^c_i-{\mathbf{s}}^{k}_i\|,\quad i=1,\ldots,N.$$
%\begin{equation}

\section{Numerical results}
\label{sec:numres}

The numerical method described in Section \ref{sec:nummet} applied to the model problems described in Section \ref{sec:M} is implemented in {\textsc{Matlab}}. In all experiments, we start by scaling the original problem such that $s_{\max}=1$ and time runs forward in the PDE. After the integration, the solution is transformed back to the original problem. 

The linear system defined in (\ref{eq:lineq}) is solved using GMRES \cite{saad1986gmres}, with an incomplete LU factorization as the preconditioner using {\tt{nofill}}. The convergence tolerance for the iterations is set to $10^{-8}$, and as the initial condition for each iteration we use the computed solution from the previous time step. 

The numerical experiments are performed on a laptop equipped with a 2.3 GHz Intel Core i7 CPU and 16 GB of RAM. The computation of the RBF-FD weights is performed in parallel using the parallel toolbox command {\tt{parfor}} with four workers.

\begin{figure}[H]
\centering
\input{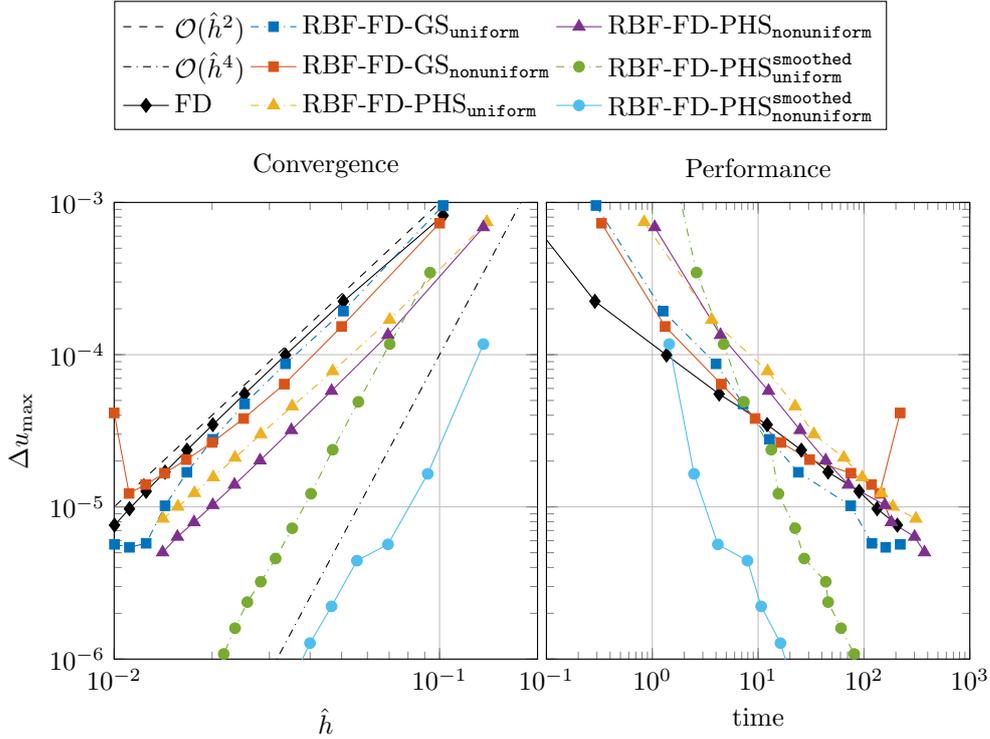}
% This file was created by matlab2tikz.
%
%The latest updates can be retrieved from
%  http://www.mathworks.com/matlabcentral/fileexchange/22022-matlab2tikz-matlab2tikz
%where you can also make suggestions and rate matlab2tikz.
%
\definecolor{mycolor1}{rgb}{0.00000,0.44700,0.74100}%
\definecolor{mycolor2}{rgb}{0.85000,0.32500,0.09800}%
\definecolor{mycolor3}{rgb}{0.92900,0.69400,0.12500}%
\definecolor{mycolor4}{rgb}{0.49400,0.18400,0.55600}%
\definecolor{mycolor5}{rgb}{0.46600,0.67400,0.18800}%
\definecolor{mycolor6}{rgb}{0.30100,0.74500,0.93300}%
\begin{tikzpicture}[trim axis left, trim axis right, baseline]

  \begin{axis}[
  grid=major,
  width=0.46\textwidth,
  height=0.5\textwidth,
  at={(0\textwidth,0\textwidth)},
  scale only axis,
  unbounded coords=jump,
  xmode=log,
  xmin=1e-01,
  xmax=1000,
  xlabel={time},
  ymode=log,
  ymin=1e-06,
  ymax=1e-03,
  yticklabels={,,}, %hides y ticks
  yminorticks=true,
  ytick distance=10^1,
  xminorticks=true,
  xmajorgrids,
  % xminorgrids,
  ymajorgrids,
  % yminorgrids,
  %ylabel={$\Delta u$},
  axis background/.style={fill=white},
  title={Performance},
  legend pos=north east,
  legend style={legend cell align=left,align=left,draw=white!15!black}
  ]
  \addplot [color=black, mark=diamond*, mark options={scale = 1.3, solid, black}]
    table[row sep=crcr]{%
    0.06582786	0.000821697635051559\\
    % 0.098248712	0.000390740290111056\\
    0.287278344	0.000224138718419901\\
    % 0.653154694	0.000143644202452091\\
    1.363033105	9.94080377643199e-05\\
    % 2.475437018	7.2569263399329e-05\\
    4.280939024	5.51237913711251e-05\\
    % 7.33640343	4.31932136169522e-05\\
    12.175330572	3.46796687033725e-05\\
    % 16.521671284	2.83557808533014e-05\\
    25.616239575	2.35866771239822e-05\\
    % 32.093849714	1.98552890761392e-05\\
    45.866742939	1.69486755397803e-05\\
    % 62.758118074	1.46040730994315e-05\\
    89.876309784	1.26596809826227e-05\\
    % 105.144092832	1.10702521652658e-05\\
    133.115589004	9.72228147643741e-06\\
    % 165.803878311	8.57895159101559e-06\\
    206.569398646	7.5886574535161e-06\\
  };
  \addlegendentry{fd2}

  \addplot [color=mycolor1, dashdotted, mark=square*, mark options={scale = 0.9,solid, mycolor1}]
    table[row sep=crcr]{%
    0.294333181	0.000954288395106928\\
    % 0.670892027	0.000392330278500077\\
    1.264234185	0.000193301090642285\\
    % 2.320541893	0.000122420681927398\\
    4.002661134	8.70798482893939e-05\\
    % 4.994922184	6.36334257922318e-05\\
    7.21062104	4.74942057093787e-05\\
    % 18.76967603	3.60763660614653e-05\\
    12.789741194	2.78000505424987e-05\\
    % 32.427884469	2.16678920812105e-05\\
    23.987790138	1.68818812815268e-05\\
    % 57.771913045	1.31270985772661e-05\\
    74.876368296	1.01673887758155e-05\\
    % 95.476722184	7.74333548296188e-06\\
    118.737986566	5.76537132542312e-06\\
    % 76.838053609	5.30633081058385e-06\\
    160.362616102	5.4245897675026e-06\\
    % 110.767437067	5.55102991257883e-06\\
    220.215181191	5.67489189767789e-06\\
  };
  \addlegendentry{gs reg}

  \addplot [color=mycolor2, mark=square*, mark options={scale = 0.9, solid, mycolor2}]
    table[row sep=crcr]{%
    0.331154439	0.000730421951017157\\
    % 0.784126736	0.000329175837351009\\
    1.321707978	0.000153151719320656\\
    % 2.407973012	9.12091426900526e-05\\
    4.492199157	6.40359806401113e-05\\
    % 5.900866537	4.81614980887175e-05\\
    9.359380277	3.80636671498055e-05\\
    % 22.562775191	3.12860705828644e-05\\
    16.399435412	2.64971538777616e-05\\
    % 40.433655359	2.30278475184913e-05\\
    30.67468922	2.04396388880333e-05\\
    % 77.191195997	1.83902580481697e-05\\
    75.199079076	1.664986889751e-05\\
    % 138.453168475	1.52056691880287e-05\\
    117.664701023	1.40298458809716e-05\\
    % 117.269431288	1.3068540435663e-05\\
    140.823239139	1.22566457229217e-05\\
    % 175.124017293	1.24037094471896e-05\\
    219.44733059	4.14165843865555e-05\\
  };
  \addlegendentry{gs adap}

  \addplot [color=mycolor3, dashdotted, mark=triangle*, mark options={scale = 1.3,solid, mycolor3}]
    table[row sep=crcr]{%
    0.837605647	0.000743167977149496\\
  % 2.025372492	0.000310529558508282\\
  3.639769825	0.000169616596950958\\
  % 7.256226552	0.000109920965315556\\
  12.290603411	7.78916616573505e-05\\
  % 18.44024452	5.851743112463e-05\\
  22.358980677	4.57511141431395e-05\\
  % 52.540076485	3.6692234042502e-05\\
  33.609578206	3.00189176842408e-05\\
  % 80.7768459	2.49649788940307e-05\\
  64.471801581	2.10393559688313e-05\\
  % 132.364579792	1.79376384123464e-05\\
  96.176644434	1.56518944708431e-05\\
  % 215.934641416	1.38316136247436e-05\\
  145.624530492	1.23307428878117e-05\\
  % 165.50100046	1.10815043109155e-05\\
  187.614598032	1.0028502564801e-05\\
  % 219.971299033	9.1350846760721e-06\\
  309.252198757	8.36410001815724e-06\\
  };
  \addlegendentry{phs reg}

  \addplot [color=mycolor4, mark=triangle*, mark options={scale = 1.3,solid, mycolor4}]
    table[row sep=crcr]{%
    1.053425471	0.000687264699283568\\
    % 2.34448544	0.00024414135768297\\
    4.408053073	0.000134546330187221\\
    % 6.595062342	8.47688371809069e-05\\
    12.539242522	5.78232547853892e-05\\
    % 18.967676855	4.16182867751698e-05\\
    24.935753374	3.19452679388398e-05\\
    % 57.789232721	2.5275475873842e-05\\
    43.468228217	2.01883525170823e-05\\
    % 92.989813402	1.6629536537395e-05\\
    71.169584803	1.39793874041304e-05\\
    % 156.944794247	1.1851168449626e-05\\
    157.613770699	1.02810154664346e-05\\
    % 267.749507248	9.01480234062266e-06\\
    181.066750603	7.93251233059677e-06\\
    % 215.198176558	7.0885477311787e-06\\
    300.717458059	6.36362730365575e-06\\
    % 312.620094495	5.65527662367907e-06\\
    372.861778238	5.02720013236327e-06\\
  };
  \addlegendentry{phs adap}

  \addplot [color=mycolor5, dashdotted, mark=*, mark options={solid, mycolor5}]
    table[row sep=crcr]{%
    % 204.797699017	1.56216283858429e-06\\
    % 137.21974258	1.60011950256348e-06\\
    % 108.867301159	1.6449609870088e-06\\
    % 122.361616773	1.66133358668564e-06\\
    % 108.698068286	1.69119461798347e-06\\
    % 85.366205378	1.8310173203911e-06\\
    100.308480392	8.86676792243582e-07\\
    80.801123362	1.08094053489867e-06\\
    60.677620782	1.59605544365503e-06\\
    45.861328437	2.36856397064627e-06\\
    43.590460923	3.22326044284102e-06\\
    27.272527995	4.57864276742076e-06\\
    22.378082203	7.23881122826828e-06\\
    15.556166938	1.22034416352584e-05\\
    13.325419311	2.37328354763949e-05\\
    7.335637489	4.89880039680132e-05\\
    4.719910087	0.000117282713859648\\
    2.619557301	0.000346118628974414\\
    1.731366747	0.00126121370849494\\
  };
  \addlegendentry{phs reg smoothed}

  \addplot [color=mycolor6, mark=*, mark options={solid, mycolor6}]
    table[row sep=crcr]{%
    1.444674318	0.000117373463206207\\
    2.478047665	1.64849648476599e-05\\
    4.160072468	5.66874455900854e-06\\
    7.925728684	4.43330995397034e-06\\
    10.689505454	2.22068321180727e-06\\
    16.261363834	1.27354519761577e-06\\
    21.123572908	6.99972089132939e-07\\
    % 24.955379512	1.04388269646372e-06\\
    % 42.464244356	1.15139006272302e-06\\
    % 45.719717467	1.21859483323073e-06\\
    % 64.871214182	1.27935297984075e-06\\
    % 84.073613238	1.35229464069946e-06\\
    % 109.512552142	1.39141329373507e-06\\
    % 115.041296553	1.36198724156136e-06\\
    % 122.191878887	1.41872953168874e-06\\
    % 137.262390987	1.40618983384723e-06\\
    % 156.076144824	1.40203105304789e-06\\
    % 170.189141117	1.33335907658125e-06\\
    % 173.101519462	1.3002449169284e-06\\
  };
  \addlegendentry{phs adap smoothed}
  \legend{};
\end{axis}
\end{tikzpicture}%
%\makebox[\linewidth][c]{%
%\centering
%\begin{subfigure}[H]{\textwidth}
%\centering
%\input{tikz/BSeuCallconv.tikz}
%\caption{$\Delta u_{\max}$ as a function of $\sqrt{N}$.}
%\end{subfigure} %\hspace{-3.5cm}
%\begin{subfigure}[H]{\textwidth}
%\centering
%\input{tikz/BSeuCalltime.tikz}
%\caption{$\Delta u_{\max}$ as a function of CPU-time.}
%\end{subfigure}
%}
\caption{$\Delta u_{\max}$ as a function of $\hat{h}$ and CPU-time in seconds for the European call option.}
\label{fig:res1}
\end{figure}

In Figure \ref{fig:res1} we plot the error $\Delta u_{\max}$ as a function of $\hat{h}\equiv1/\sqrt{N}$ as well as of CPU-time for the model problem. The error is defined as 
\begin{equation}
\Delta u(s_1,s_2)=|u^c(s_1,s_2,0)-u^*(s_1,s_2,0)|,
\label{eq:error}
\end{equation}
where $u^c$ is the computed solution and $u^*$ is a reference solution computed with a second order finite difference method on a very fine grid. We use (\ref{eq:error}) to define 
\begin{equation}
\Delta u_{\max}=\max_{[s_1,s_2]\in\hat{\Omega}}\Delta u(s_1,s_2),
\label{eq:max}
\end{equation}
where $\hat{\Omega}=\left[\frac{1}{3}K,\frac{5}{3}K\right]\times \left[\frac{1}{3}K,\frac{5}{3}K\right]$. %In Table \ref{tab:space}, we display the spatial discretizations used in Figure \ref{fig:res1}.%
We denote standard second order finite differences \cite{tavella2000pricing} by FD. RBF-FD-GS is an RBF-FD method with Gaussian RBFs, stencil size $m=25$ and a node density dependent shape parameter that is presented in detail in \cite{milovanovic2018radial}. Abbreviation RBF-FD-PHS is used for the method that is presented in this paper. Moreover, we use designation \texttt{smoothed} in the superscript for the computations performed with the smoothing of the initial data, and \texttt{uniform} and \texttt{nonuniform} to specify the node layouts.
%\begin{table}[htb]
%\centering
%\begin{tabular}{|l|l|l|l|} \hline
%\textbf{method} & \textbf{spatial} & \textbf{node} & \textbf{final}  \\ 
%&\textbf{discretization}&\textbf{layout}&\textbf{condition} \\ \hline
%FD&2nd order FD&Uniform&Original\\
%RBF-FD-GS$_{\mathrm{uniform}}$&RBF-FD, Gauss. RBFs&Uniform&Original\\
%RBF-FD-GS$_{\mathrm{nonuniform}}$&RBF-FD, Gauss. RBFs&Nonniform&Original\\
%RBF-FD-PHS$_{\mathrm{uniform}}$&RBF-FD, PHS RBFs&Uniform&Original\\
%RBF-FD-PHS$_{\mathrm{nonuniform}}$&RBF-FD, PHS RBFs&Nonniform&Original\\
%RBF-FD-PHS$_{\mathrm{uniform}}^{\mathrm{smoothed}}$&RBF-FD, PHS RBFs&Uniform&Smoothed\\
%RBF-FD-PHS$_{\mathrm{nonuniform}}^{\mathrm{smoothed}}$&RBF-FD, PHS RBFs&Nonniform&Smoothed\\
%\hline
%\end{tabular}
%\caption{Spatial discretizations used in Figure \ref{fig:res1}.}
%\label{tab:space}
%\end{table}

Independent of spatial discretization that is used, we employ BDF2 with $M=N_s$ time steps in all experiments. For the RBF-FD methods $N_s$ is defined in (\ref{eq:Ns}) and for FD it is defined by $N_s=\sqrt{N}$. With this number of time steps, the temporal discretization error is not visible in the plots. 

%\begin{figure}[H]
%\centering
%\makebox[\linewidth][c]{%
%\centering
%\begin{subfigure}[H]{0.8\textwidth}
%\centering
%\input{tikz/grid2dreg.tikz}
%\caption{$\Delta u_{\max}$ as a function of $\sqrt{N}$.}
%\end{subfigure} \hspace{-3.5cm}
%\begin{subfigure}[H]{0.8\textwidth}
%\centering
%\input{tikz/grid2dadap.tikz}
%\caption{$\Delta u_{\max}$ as a function of CPU-time.}
%\end{subfigure}
%}
%\caption{$\Delta u_{\max}$ as a function of $\sqrt{N}$ and CPU-time for the American put option.}
%\label{fig:res2}
%\end{figure}

In Figure \ref{fig:res1} we see that all methods, but the two that are using a smoothed final condition, exhibit second order convergence. Among those five second order methods, RBF-FD with PHS exhibits the smallest error for a given $N$, and a nonuniform node layout gives a smaller error than the uniform one using the same $N$. RBF-FD-PHS with a smoothed final condition exhibits fourth order spatial convergence whether we are using a uniform or nonuniform node layout (apart from a small deviation for the nonuniform node layout). When it comes to computational time to reach a certain $\Delta u_{\max}$, FD is competitive for the larger errors displayed. This makes sense since the RBF-FD methods all have to compute the weights $w_k$, $k=1,\ldots,m$ before the time-stepping. Moreover, our model problem has a fairly short time to maturity $T=0.2$. For longer times to maturity, FD does not perform equally well compared to the RBF methods, see \cite{milovanovic2018radial, Slobodan}. We also establish that the fourth order methods quickly become superior when it comes to CPU-time to reach a certain $\Delta u_{\max}$. That is especially true for RBF-FD-PHS$_{\mathrm{nonuniform}}^{\mathrm{smoothed}}$. Even though this method has a computational prephase that includes both computation of weights $w_k$, $k=1,\ldots,m$ and smoothing of the final condition, the method requires a much smaller CPU-time than the other methods for $\Delta u_{\max}<10^{-4}$.

\begin{figure}[H]
\centering
\begin{tikzpicture}[trim axis left, trim axis right,baseline]
    \begin{axis}[
        axis on top,
        grid=major,
        width=1.6*0.3515\textwidth,
        height=1.6*0.3515\textwidth,
        % scale only axis,
        % enlargelimits=false,
        % xmode=log,
        xmin=0,
        xmax=4,
        % ymode=log,
        ymin=0,
        ymax=4,
        yminorticks=true,
        xminorticks=true,
        xlabel={$s_1$},
        ylabel={$s_2$},
        y label style={at={(+0.1,0.5)}},
        % ytick distance=10^1,
        title={$\text{RBF-FD-PHS}_{\texttt{uniform}}$},
        ]

      \addplot graphics[xmin=0,ymin=0,xmax=4,ymax=4] {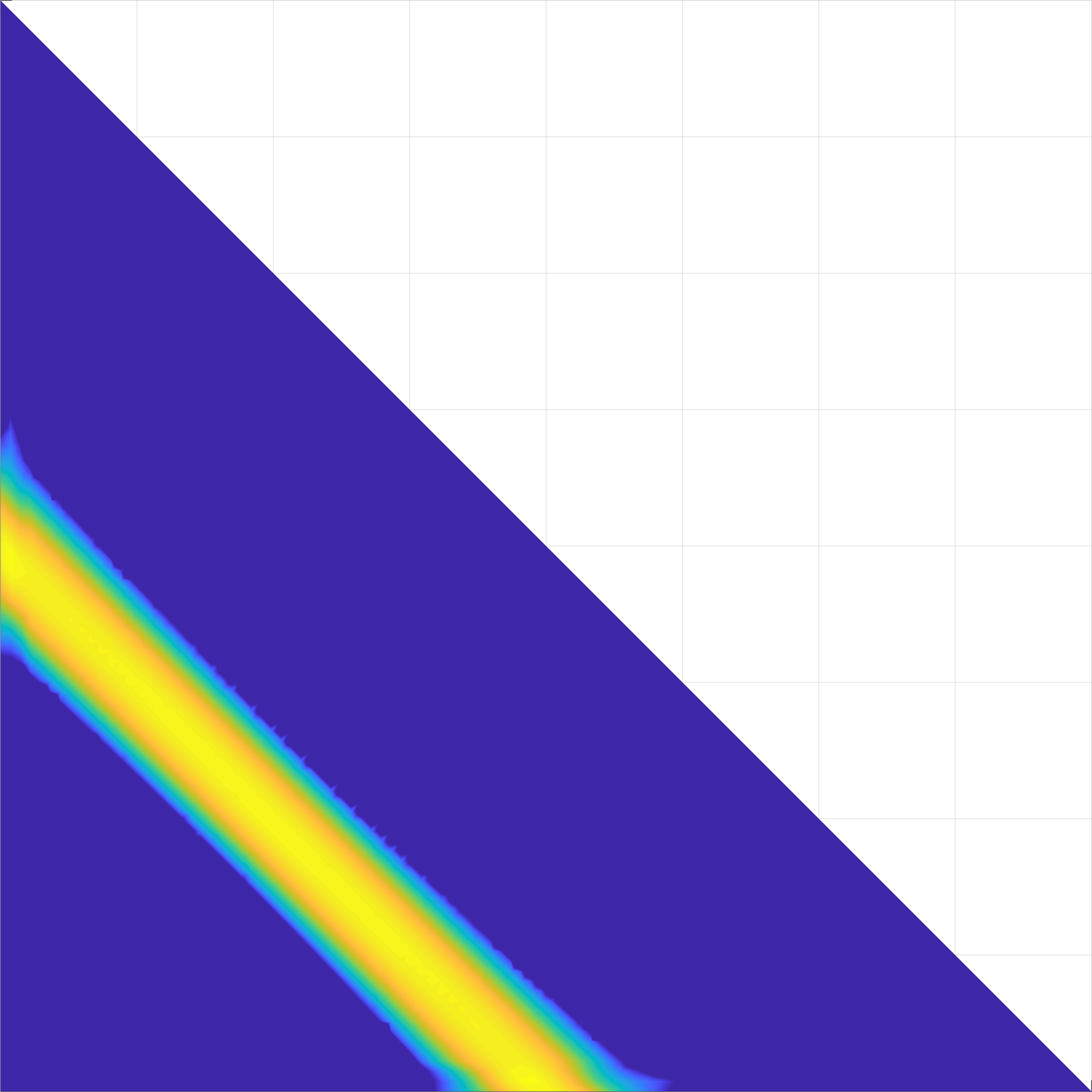};
      \draw [white, dashdotted, thick] (axis cs:1/3,1/3) rectangle (axis cs:5/3,5/3);
      %  \addplot[
      %   black,
      %   domain=0:4,
      %   samples=100,
      %   ]
      %   {0.0015/x} [every node/.style={yshift=7pt},sloped]
      %         node[pos=0.75] {$\varepsilon^*(h)$}
      %         ;
    \end{axis}
  \end{tikzpicture} \hspace{-0.1cm}
\begin{tikzpicture}[trim axis left, trim axis right,baseline]
    \begin{axis}[
        axis on top,
        grid=major,
        width=1.6*0.3515\textwidth,
        height=1.6*0.3515\textwidth,
        % scale only axis,
        % enlargelimits=false,
        % xmode=log,
        xmin=0,
        xmax=4,
        % ymode=log,
        ymin=0,
        ymax=4,
        yticklabels={,,},
        yminorticks=true,
        xminorticks=true,
        xlabel={$s_1$},
        % ylabel={$s_2$},
        % ytick distance=10^1,
        title={$\text{RBF-FD-PHS}^{\texttt{smoothed}}_{\texttt{uniform}}$},
        ]

      \addplot graphics[xmin=0,ymin=0,xmax=4,ymax=4] {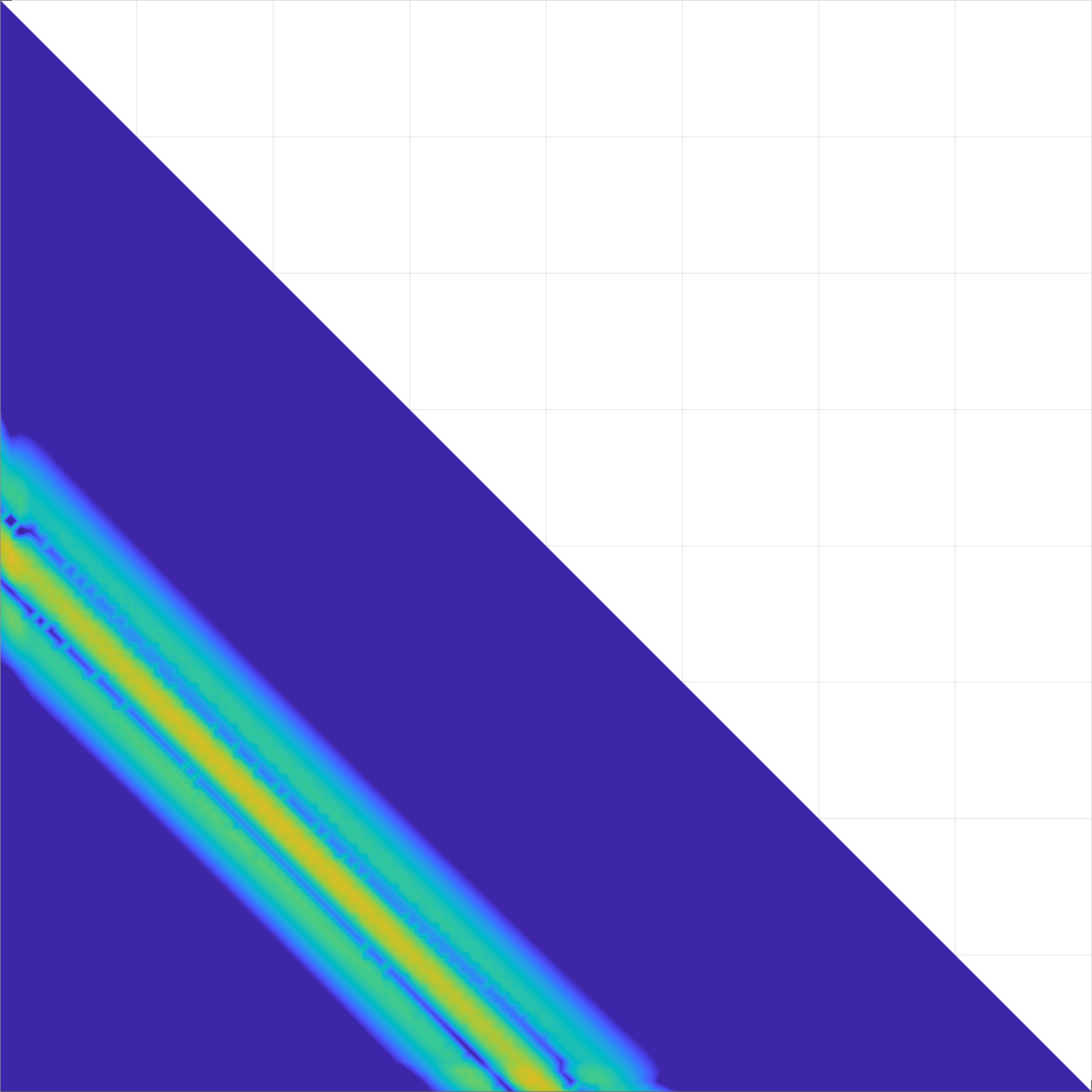};

      \draw [white, dashdotted, thick] (axis cs:1/3,1/3) rectangle (axis cs:5/3,5/3);

      %  \addplot[
      %   black,
      %   domain=0:4,
      %   samples=100,
      %   ]
      %   {0.0015/x} [every node/.style={yshift=7pt},sloped]
      %         node[pos=0.75] {$\varepsilon^*(h)$}
      %         ;
    \end{axis}
  \end{tikzpicture} \hspace{-0.1cm}
\begin{tikzpicture}[baseline, trim axis left, trim axis right,baseline]
\pgfplotsset{xtick style={draw=none}, every axis y label/.append style={yshift=0cm}}
    \begin{axis}[
        axis on top,
        width=4*0.0070\textwidth,
        height=0.43197 \textwidth,
        scale only axis,
        enlargelimits=false,
        xmin=0,
        xmax=1,
        ymode=log,
        ymin=1e-06,
        ymax=1e-04,
        % ytick distance=10^1,
        ytick={1e-06,1e-05,1e-04},
        extra y ticks={1e-02},
        extra y tick labels={$\geq10^{-2}$},
        %yminorticks=true,
        title={$\Delta u$},
        ylabel near ticks, yticklabel pos=right,
        xticklabels={,,},
        ]

      \addplot graphics[xmin=0,ymin=1e-06,xmax=1,ymax=1e-04]{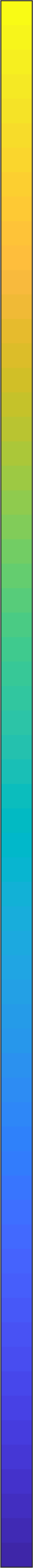};
    \end{axis}
  \end{tikzpicture}
%
%\makebox[\linewidth][c]{%
%\centering
%\begin{subfigure}[H]{0.8\textwidth}
%\centering
%\input{tikz/grid2dreg.tikz}
%\caption{$\Delta u$ for the uniform node layout.}
%\end{subfigure} \hspace{-3.5cm}
%\begin{subfigure}[H]{0.8\textwidth}
%\centering
%\input{tikz/grid2dadap.tikz}
%\caption{$\Delta u$ for the nonuniform node layout.}
%\end{subfigure}
%}
\caption{Heat maps of $\Delta u$ for the European call basket option on uniform node layouts. The boundary of $\hat{\Omega}$ is marked with a white dash-dotted line.}
\label{fig:res3}
\end{figure}

\begin{figure}[H]
\centering
\begin{tikzpicture}[trim axis left, trim axis right,baseline]
    \begin{axis}[
        axis on top,
        grid=major,
        width=1.6*0.3515\textwidth,
        height=1.6*0.3515\textwidth,
        % scale only axis,
        % enlargelimits=false,
        % xmode=log,
        xmin=0,
        xmax=4,
        % ymode=log,
        ymin=0,
        ymax=4,
        yminorticks=true,
        xminorticks=true,
        xlabel={$s_1$},
        ylabel={$s_2$},
        y label style={at={(+0.1,0.5)}},
        % ytick distance=10^1,
        title={$\text{RBF-FD-PHS}_{\texttt{nonuniform}}$},
        ]

      \addplot graphics[xmin=0,ymin=0,xmax=4,ymax=4] {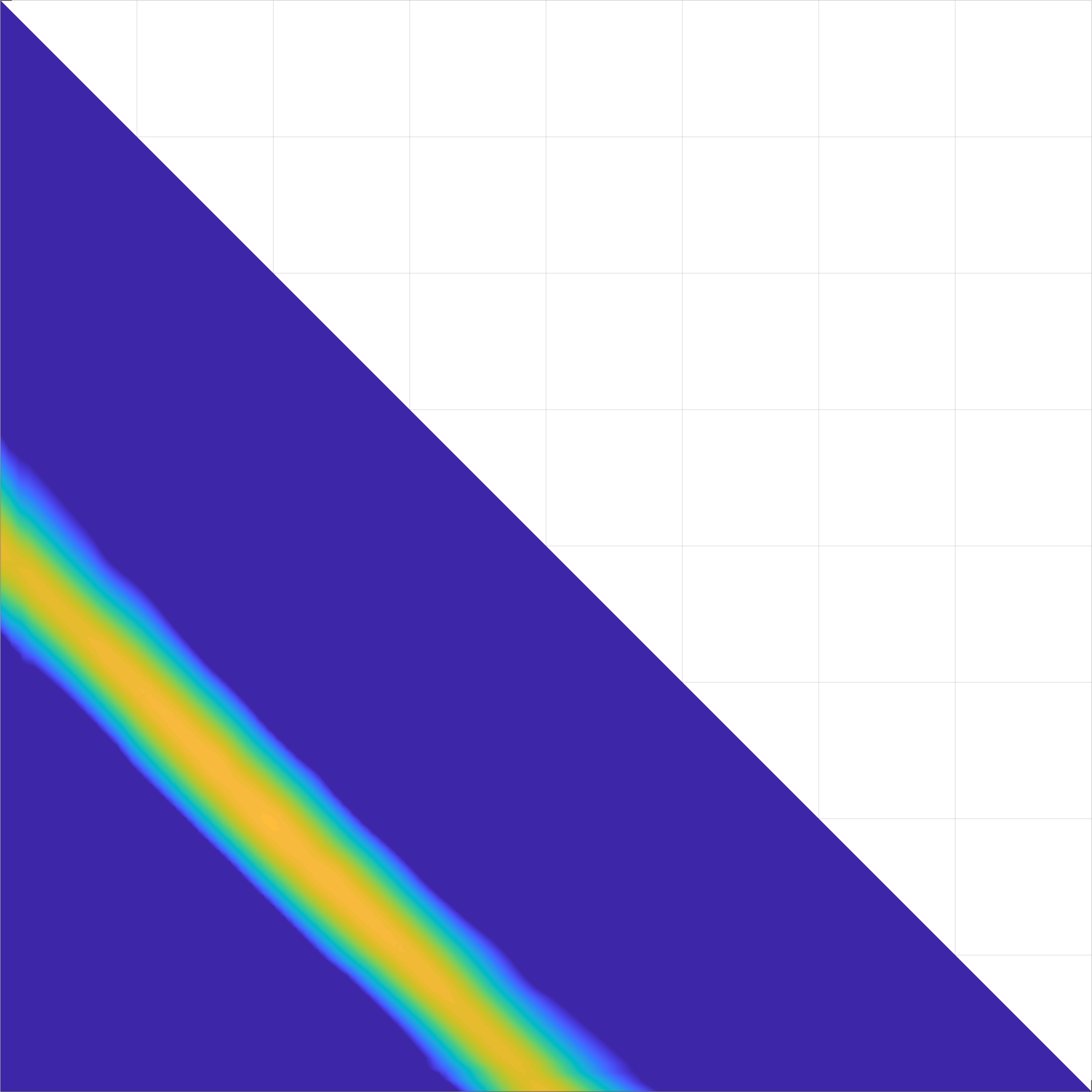};
      \draw [white, dashdotted, thick] (axis cs:1/3,1/3) rectangle (axis cs:5/3,5/3);
      %  \addplot[
      %   black,
      %   domain=0:4,
      %   samples=100,
      %   ]
      %   {0.0015/x} [every node/.style={yshift=7pt},sloped]
      %         node[pos=0.75] {$\varepsilon^*(h)$}
      %         ;
    \end{axis}
  \end{tikzpicture} \hspace{-0.1cm}
\begin{tikzpicture}[trim axis left, trim axis right,baseline]
    \begin{axis}[
        axis on top,
        grid=major,
        width=1.6*0.3515\textwidth,
        height=1.6*0.3515\textwidth,
        % scale only axis,
        % enlargelimits=false,
        % xmode=log,
        xmin=0,
        xmax=4,
        % ymode=log,
        ymin=0,
        ymax=4,
        yticklabels={,,},
        yminorticks=true,
        xminorticks=true,
        xlabel={$s_1$},
        % ylabel={$s_2$},
        % ytick distance=10^1,
        title={$\text{RBF-FD-PHS}^{\texttt{smoothed}}_{\texttt{nonuniform}}$},
        ]

      \addplot graphics[xmin=0,ymin=0,xmax=4,ymax=4] {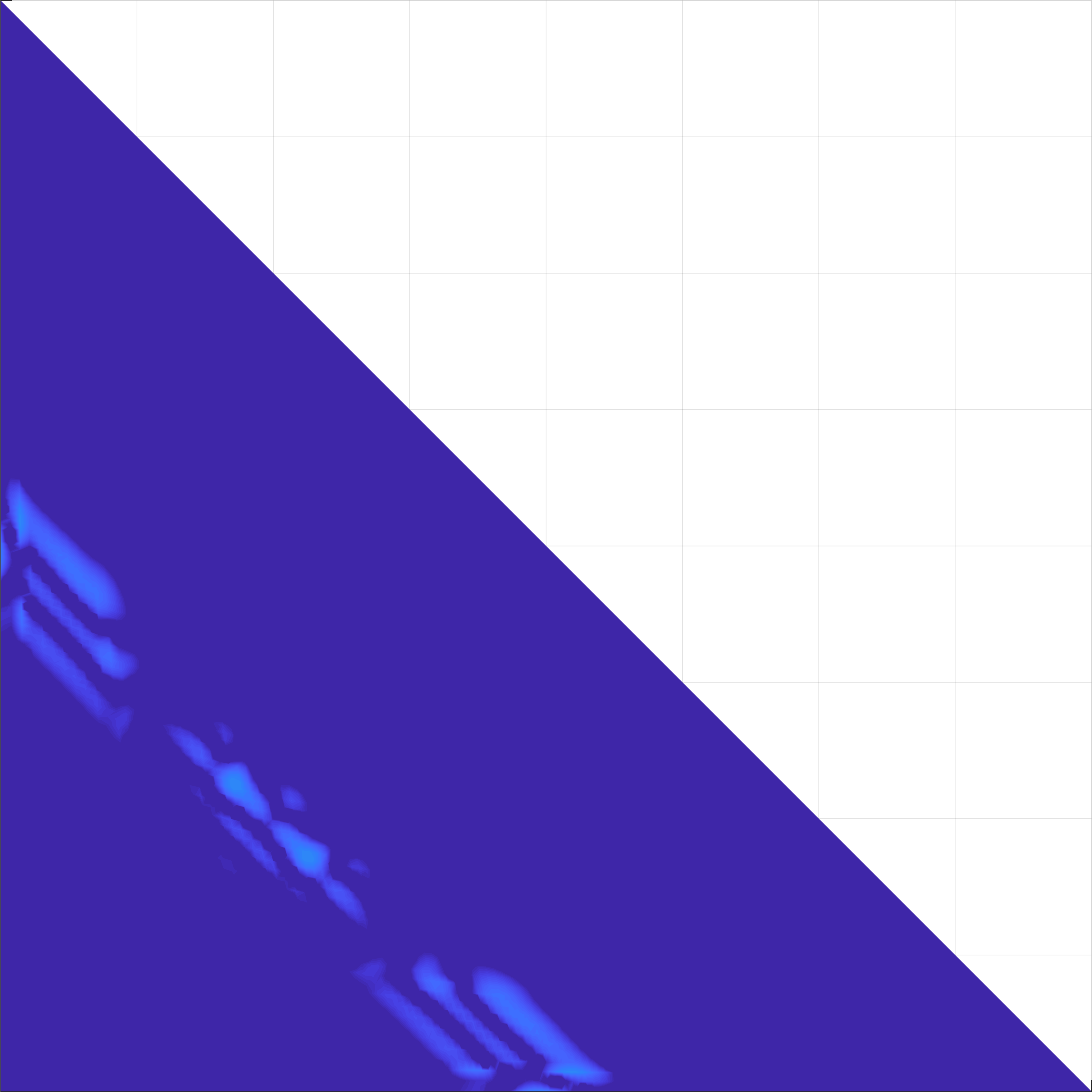};

      \draw [white, dashdotted, thick] (axis cs:1/3,1/3) rectangle (axis cs:5/3,5/3);

      %  \addplot[
      %   black,
      %   domain=0:4,
      %   samples=100,
      %   ]
      %   {0.0015/x} [every node/.style={yshift=7pt},sloped]
      %         node[pos=0.75] {$\varepsilon^*(h)$}
      %         ;
    \end{axis}
  \end{tikzpicture} \hspace{-0.1cm}
\begin{tikzpicture}[baseline, trim axis left, trim axis right,baseline]
\pgfplotsset{xtick style={draw=none}, every axis y label/.append style={yshift=0cm}}
    \begin{axis}[
        axis on top,
        width=4*0.0070\textwidth,
        height=0.43197 \textwidth,
        scale only axis,
        enlargelimits=false,
        xmin=0,
        xmax=1,
        ymode=log,
        ymin=1e-06,
        ymax=1e-04,
        % ytick distance=10^1,
        ytick={1e-06,1e-05,1e-04},
        extra y ticks={1e-02},
        extra y tick labels={$\geq10^{-2}$},
        %yminorticks=true,
        title={$\Delta u$},
        ylabel near ticks, yticklabel pos=right,
        xticklabels={,,},
        ]

      \addplot graphics[xmin=0,ymin=1e-06,xmax=1,ymax=1e-04]{tikz/BSeuCall_phs_colorbar.png};
    \end{axis}
  \end{tikzpicture}
%\makebox[\linewidth][c]{%
%\centering
%\begin{subfigure}[H]{0.8\textwidth}
%\centering
%\input{tikz/grid2dreg.tikz}
%\caption{$\Delta u$ for the uniform node layout.}
%\end{subfigure} \hspace{-3.5cm}
%\begin{subfigure}[H]{0.8\textwidth}
%\centering
%\input{tikz/grid2dadap.tikz}
%\caption{$\Delta u$ for the nonuniform node layout.}
%\end{subfigure}
%}
\caption{Heat maps of $\Delta u$ for the European call basket option on nonuniform node layouts. The boundary of $\hat{\Omega}$ is marked with a white dash-dotted line.}
\label{fig:res4}
\end{figure}
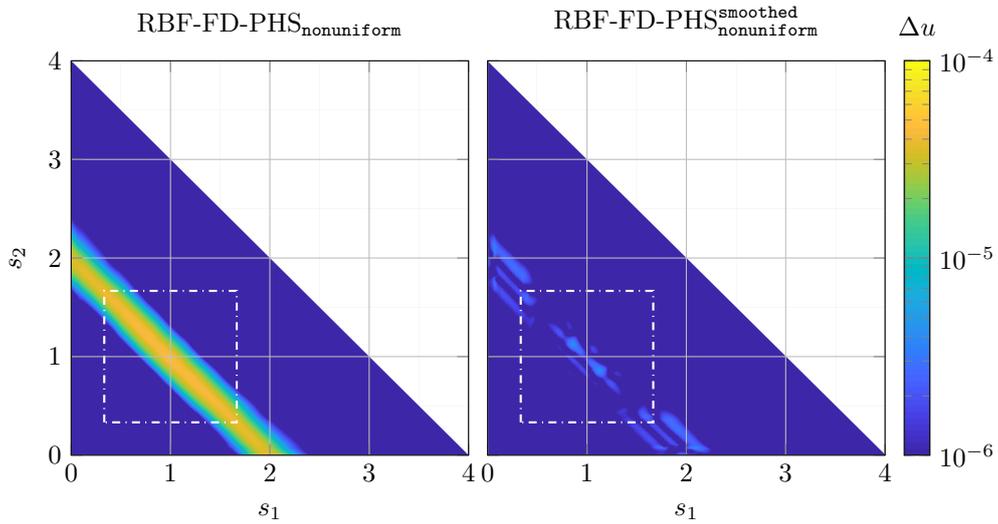

In Figures \ref{fig:res3}--\ref{fig:res4} we show a heat map of the error $\Delta u$ defined in (\ref{eq:error}) for the model problem using the method RBF-FD-PHS. In Figure \ref{fig:res3} we present the error on the uniform node layout, and in Figure \ref{fig:res4} the error on the nonuniform node layout, with $N=6105$ for both node layouts. To the right in both figures, we have used the smoothed final condition, while the original one is used in the plots on the left. The errors in Figures \ref{fig:res3}--\ref{fig:res4} are presented for $0\le s_j\le 4$, $j=1,2$, in order to have a better view of the error profile around the smoothed area. The color scale is the same in all four plots. 

From Figures \ref{fig:res3}--\ref{fig:res4} we conclude that the smoothing of the final condition renders a $\Delta u$ that has a smaller magnitude compared to the original final condition. Moreover, $\Delta u$ obtained from the smoothed final condition has three maxima along a line $s_2=s_1+const.$ while the corresponding number of maxima is 1 for the nonsmoothed final condition. We also note that the magnitude of  $\Delta u$ is smaller for the nonuniform node layout compared to the uniform one. That is due to the fact that for the nonuniform node layout, the number of nodes is larger in the area where the solution has large derivatives, i.e., around the strike price. We end this section by concluding that in this particular example, $\Delta u_{\max}$ is more than one order smaller using smoothing of the final condition on a nonuniform grid than without the smoothing on a uniform grid for the same number of nodes.

\section{Conclusions}
\label{sec:concl}
In this paper, we have implemented a solver to price financial derivatives based on RBF-FD discretization in space and BDF2 in time. As RBFs we use PHSs, augmented with monomials of up to degree $p$. The formal order of this spatial discretization is $p$, however for many pricing problems the lack of smoothness of the initial data limits the actual order obtained in numerical simulations. However, by employing a smoothing technique to the initial data, the formal order of the discretization is retained. 

The RBF-FD discretizations have the advantage over standard FD such that the nodes do not have to be organized in a Cartesian grid. On the other hand, the RBF-FD discretizations have the merit that they render sparse differentiation matrices as opposed to global RBF approximations that lead to full matrices. Thus, RBF-FD has the possibility to give accurate solutions on nonuniform node layouts, still yielding sparse matrices. 

As a model problem, we consider pricing of a European type basket option issued on two underlying assets, resulting in a PDE in two spatial dimensions and time. By employing a nonuniform node layout that has a denser node distribution where we are most interested in having an accurate solution, together with smoothing of the final condition, the numerical experiments demonstrate that our developed method gives a very accurate solution in a short time using fewer nodes than the methods that we compare with, for this model problem. The fact that we can solve the problem accurately with fewer nodes becomes extremely important when we want to solve problems in higher dimensions, e.g., for pricing of financial derivatives issued on several underlying assets. Since the number of degrees of freedom grows exponentially in the number of dimensions (number of underlying assets), the ability to use fewer nodes per dimension to reach a certain accuracy might lead to the possibility to solve problems that would not be possible to solve with traditional techniques.

%\begin{figure}[H]
%%\centering
%\scalebox{0.9}{\input{tikz/SABReuCallI.tikz}}
%\caption{Results for the European call option under the SABR model, Parameter Set I. The reference values for $K_i=S_0{\rm{exp}}(0.1\times \sqrt{T} \times \delta_i)$, $\delta_i=-1.0,0.0,1.0$ are given by 0.221383196830866, 0.193836689413803, and 0.166240814653231.}
%\label{fig:SABReuCallI}
%\end{figure}

\bibliographystyle{abbrvnat}
\bibliography{high-order}

\end{document}